\documentstyle[aps,preprint,prc]{revtex}
\begin{document}

\begin{center}
\large\bf
  CHAOS AND ORDER IN THE SHELL MODEL EIGENVECTORS\\

\vspace{5mm}
\normalsize
  Vladimir Zelevinsky$^{1,2}$, Mihai Horoi$^{1,3}$ and B. Alex Brown$^{1}$\\

\vspace{.8cm}

{\em
$^{1}$National Superconducting Cyclotron Laboratory,
East Lansing, MI 48824\\
$^{2}$Budker Institute of Nuclear Physics, Novosibirsk 630090, Russia\\
$^{3}$Institute of Atomic Physics, Bucharest, Romania\\}

\end{center}

\begin{abstract}
The energies and wave functions of stationary many-body states are analyzed
to look for the signatures of quantum chaos. Shell model calculations
with the Wildenthal interaction are performed in the $J-T$ scheme
for 12 particles in the $sd$-shell.
The local level statistics are in perfect agreement with
the GOE predictions. The analysis of the amplitudes of the eigenvectors
in the shell model basis with the aid of the informational entropy and
moments of the distribution function shows evidence for local
chaos with a localization length reaching 90\% of the total dimension
in the middle of the spectrum. The degree
of chaoticity is sensitive to the the strength
of the residual interaction as compared
to the single particle energy spacing.

\vspace*{.7cm}

{\bf{PACS numbers:}} 24.60.-k, 24.60.Lz, 21.10.-k, 21.60.Cs

\end{abstract}

\newpage

Quantum chaos in many-body systems was studied mostly from the
viewpoint of level statistics which displays a clear relation to
the notion of classical chaos \cite{Haake}. Presumably much more information
could be obtained from an analysis of the wave functions and
transition amplitudes. Here one expects to encounter the transition
from the simple picture of almost independent elementary excitations
to extremely mixed compound states which would display new
specific features as, for example, so called  dynamic enhancement
of weak interactions \cite{SF}.

To perform such an analysis and to check various hypotheses
concerning complicated quantum dynamics, one needs a rich set of
data which would allow one to make statistically reliable conclusions.
Realistic nuclear shell model calculations are one of the most promising
candidates for studying this largely unknown structure of quantum
chaotic states.

We studied the behavior of the basis-state amplitudes of the shell
model eigenvectors
produced in the $J-T$ scheme for 12 particles in the $sd$
shell.
Our model hamiltonian describing a many-body system of
valence particles within a major shell contains a
one-body part, which is due to an existing core (e.g. $^{16}$O for
the $sd$ shell) and a
two-body antisymmetrized interaction of the valence particles
\begin{equation}
H = \sum \epsilon_{\mu} a^{\dagger}_{\mu} a_{\mu} +
\frac{1}{4} \sum V_{\mu \nu \lambda \rho}
a^{\dagger}_{\mu} a^{\dagger}_{\nu}
a_{\lambda} a_{\rho}\ .
\label{eq:ham}
\end{equation}

\noindent
In our calculations the Wildenthal interaction along with the
well known procedure to project
out of the $m$-scheme the states with correct values of the total angular
momentum $J$ and isospin $T$ were utilized \cite{Wild,OXBA}.
The $J-T$ projected states $\mid k\rangle$ are used to build the matrix
of the many-body hamiltonian,
$H_{k k'} = \langle J T;\ k \mid H \mid J T;\ k' \rangle$,
which is eventually diagonalized producing the eigenvalues $E_{\alpha}$
and the eigenvectors
\begin{equation}
\mid J T ;\ \alpha \rangle = \sum_{k} C_{k}^{\alpha} \mid J T ;\ k\rangle .
\label{eq:eigv}
\end{equation}
\noindent
They represent the object of our investigation.

The matrix dimension for the $J^{\pi}T = 2^{+}0$ states is 3273. The
density of states steeply increases along with excitation energy, reaches
its maximum and then decreases again for the highest energy. This high-energy
behavior, as well as the approximate symmetry with respect to the middle
of the spectrum, are artificial features of models with finite Hilbert
space in contrast to actual many-body systems. For the analysis of the
level statistics we used levels 200 - 3000.

Fig. 1 shows the standard quantities which define the chaoticity
of a quantum system \cite{Haake,Brody}, the unfolded distribution
of the nearest neighbor spacings $P(s)$
and the spectral rigidity $\Delta_{3}$,
for this class of states. The solid lines in
both parts of the figure describe the random matrix results for the Gaussian
Orthogonal Ensemble (GOE). The dashed line on the right corresponds to
the Poisson level distribution which is characteristic of
an ordered system. The closeness of $\Delta_{3}$ to
the random matrix results even for very large values of $L$ is remarkable.
Previous to this study the largest value of $L$ considered was 80 \cite{Arve}.
Thus, the level statistics manifest generic chaotic behavior.

We next look to the structure of the wave functions
which could reveal in more detail how close to chaoticity we are.
The appropriate quantities to measure the degree of complexity of a given
eigenstate $|\alpha\rangle$, eq.(2), with respect to the original shell
model basis are,
for instance, the informational entropy \cite{Izr,Reichl},
\begin{equation}
S^{\alpha} = - \sum_{k}\mid C^{\alpha}_{k} \mid^{2}
\ln \mid C^{\alpha}_{k} \mid^{2},
\label{eq:end}
\end{equation}
\noindent
or the moments of the distribution of amplitudes $|C^{\alpha}_{k}|^{2}$.
The second
moment determines the number of principal components ($NPC$) of an eigenvector
$|\alpha \rangle$,
\begin{equation}
(NPC)^{\alpha} = \left(\sum_{k}\mid C^{\alpha}_{k} \mid^{4}
 \right)^{-1}.
\label{eq:npc}
\end{equation}
\noindent
In the GOE all basis states are completely mixed
so that the resulting eigenvectors are totally delocalized
and cover uniformly the $N$-dimensional sphere of radius 1
\cite{Brody,Perc,Berry}. Gaussian fluctuations
with zero mean, $\overline{C^{\alpha}_{k}}=0$, and width
$\overline{|C^{\alpha}_{k}|^{2}} = 1/N$ lead to the values $\ln (0.48N)$
and $N/3$ for the quantities (3) and (4) respectively. Here $N$ is
the total dimension of the model space. In reality, the
incomplete mixing of basis states determined by specific properties
of the hamiltonian
can coexist with the GOE-type level correlations.

The left upper part of Fig. 2 presents the $\exp (S^{\alpha})$ quantity
for the $2^{+}0$ states. On the $x$-axis are the
eigenstates numbered in order of their energies. This simple
"numbered" scale is equivalent to the "unfolding"
procedure described for example by Brody {\sl et al} \cite{Brody}. "Unfolding"
is introduced to separate local correlations and fluctuations from
the global spectral properties. The
solid line represents the GOE result ($0.48 N$). One observes a
semicircle-type behavior and a 12\%  deviation from the GOE
even for the maximum entropy in the middle of the spectrum.

It is interesting to study the role of single-particle energies (see
eq. (1)) for the chaotic behavior of the amplitudes. The upper right part of
Fig. 2 shows the $\exp(S^{\alpha})$ quantity for the $2^{+}0$ states
for the same hamiltonian but with all
single particle energies, $\epsilon_{\mu}$ in eq. (1), set to zero.
In this degenerate case the GOE limit (solid line) is attained and
the chaotic regime extends over a larger part of the spectrum.

To further quantify these effects,
we look to the distribution $P(l_{S})$ of
$l_{S}^{\alpha} = \exp (S^{\alpha}) / 0.48 N$.
One can interprete $l_{S}^{\alpha}$ as a delocalization length $N^{\alpha}/N$.
It is expected to be shaped around $l_S = 1$ in the chaotic limit.
The lower left panel of Fig. 3  presents the results for
the values of $l_{S}$ calculated for the normal hamiltonian and shown
on the upper left panel. Here the limit of $l_S = 1$ is not
reached. On the other hand, for degenerate single-particle orbitals
(upper and lower panels on the right side), the distribution of
localization lengths is more narrow and
the full chaotic limit is reached. This is related to the fact that
the mean field in general tends to smooth out the chaotic aspects of many-body
dynamics \cite{Zel93}.

The number of principal components (4) behaves in a very similar way gradually
increasing from the edges of the spectrum to the middle, Fig.3 (left).
Even the most complicated states are shifted down from the GOE
limit of complete mixing. However, for
the ratio $\exp S^{\alpha}/(NPC)^{\alpha}$ one obtains the results in
the right part
of Fig. 3. For a Gaussian distribution of amplitudes $C^{\alpha}_{k}$ of
a given eigenvector $|\alpha\rangle$, this ratio would be given by
the universal ($N$-independent) random
matrix result equal to 1.44 (solid line). The
flattened region indicates that the chaotic dynamics,
even if not complete, extends far beyond
the region nearby the maximum of the informational entropy. Again we use
the "unfolded" numbered scale rather than the energy scale.
The "unfolding" reveals
the presence of "local" chaos: in a given small energy range,
the eigenstates are characterized by a typical
delocalization length $N^{\alpha} < N$ and by a Gaussian distribution of the
amplitudes $C^{\alpha}_{k}$ with zero mean value and variance
$(N^{\alpha})^{-1}$. This length cancels in the ratio
$\exp (S^{\alpha}) /(NPC)^{\alpha}$ for the majority of states
in the middle of the spectrum. The flatness of this ratio, as compared
to strong $\alpha$-dependence of $\exp (S^{\alpha})$ and $(NPC)^{\alpha}$
separately, indicates the existence of the local chaotic properties
scaling with $N^{\alpha}$. The edge regions with this ratio larger than 1
clearly correspond to relatively weakly mixed states with a reduced
$NPC$. We note the very narrow dispersion of points
in Figs. 2 and 3 for our measures of complexity.

	Using the same tools we also analyzed the $J^{\pi}T = 0^{+}0$
states of the model (the dimension of this subspace is 839) and obtained
very similar results.

In conclusion, we have studied the chaotic properties of a
many-body quantum system which consists of 12 valence
particles interacting in the $sd$-shell. We have shown that standard
signatures of chaos like nearest neighbor spacing distribution or
spectral rigidity are not sensitive enough to show
deviations from the random matrix results. The
informational entropy or moments of the $\mid C^{\alpha}_{k} \mid^{2}$
distribution like $(NPC)$ are much more suited to reveal these details.
Arguments are given in favor of local chaos
characterized by a Gaussian
distribution of the components of the wave functions with the variance
related to the localization length. This length grows as the level density
increases but does not reach the GOE limit.
Finally, we have shown that the effect of the core (given by the
single-particle energies) diminishes the maximal degree of chaoticity
which can be obtained when the system consists of interacting
valence particles only.

We point out that our signatures of complexity, $l_{S}$ and ($NPC$), are
basis-dependent, they reflect mutual properties of the eigenbasis of the
problem and the original "simple" basis $|k\rangle$. In the basis of
eigenvectors of a random matrix, unrelated
to the mean field of the problem, the results have been checked to coincide
with those for the GOE as it should be due to the orthogonal invariance.
The basis dependence can give additional physical information and
it should be studied separately.
The mean field basis is in some sense
exceptional since it can be shown that the mean field itself
is generated by averaging out the most chaotic components of many-body
dynamics \cite{Zel93}. Therefore it can be considered as a
preferential representation
for our purpose.
It is remarkable that the "natural"
choice of the shell model
basis sheds a detailed light on the global and local
chaotic properties of the wave functions in the many-body system
with strong interaction.

\vspace{2cm}

The authors would like to acknowledge support from the
NSF grant 94-03666.

\newpage
\vspace{3cm}

\begin{center}
{\bf Figure captions}
\end{center}

\vspace{1.5cm}

{\bf Figure 1} \
Unfolded distribution of the nearest neighbor spacings, $P(s)$, left,
and the rigidity of the spectrum, $\Delta_{3}$, right,
for $2^{+}0$ states.\\

{\bf Figure 2} \
Left panel: exponential of entropy (upper part), and the distribution of
$l_{S} = \exp S/0.48 N$ for $2^{+}0$ states calculated with the full
hamiltonian of the model (lower part); right panel: the same quantities
for the degenerate model with $\epsilon_{\mu} = 0$.\\

{\bf Figure 3} \
The Number of Principal Components, eq.(4), of $2^{+}0$ states,
left, and the ratio $\exp S/(NPC)$ for the same states, right.

\end{document}